\documentclass[pre,aps,showpacs,twocolumn,floatfix]{revtex4}
\usepackage[dvips]{graphicx}
\usepackage{amsmath}
\usepackage{amsfonts}
\usepackage{amssymb}
\usepackage{times}

\newcommand{\be}{\begin{equation}}
\newcommand{\ee}{\end{equation}}
\newcommand{\bea}{\begin{eqnarray}}
\newcommand{\eea}{\end{eqnarray}}
\newcommand{\ba}{\begin{array}}
\newcommand{\ea}{\end{array}}

\begin{document}

%
%
\title{Resonant activation in the presence of non-equilibrated baths.}

\author{Bart{\l}omiej \surname{Dybiec}}
\email{bartek@th.if.uj.edu.pl}

\author{Ewa \surname{Gudowska--Nowak}}
\email{gudowska@th.if.uj.edu.pl}
\affiliation{Marian~Smoluchowski Institute of Physics,\\
 Jagellonian University, Reymonta~4, 30--059~Krak\'ow, Poland}

\date{\today}

\begin{abstract} 
We study the generic problem of the escape of a classical particle over 
a fluctuating barrier under the influence of non-Gaussian noise 
mimicking the effects of not-fully equilibrated bath. 
Our attention focuses on the effect of the stable noises on the
mean escape time and on the phenomenon of resonant activation (RA). 
Possible physical interpretation of the occurrence of L\'evy noises in the
system of interest is discussed
and the connection between 
the Tsallis statistics and the Fractional 
Fokker Planck Equation is addressed.
\end{abstract}


\maketitle

\section{introduction}
In a classical Langevin approach the influence of the bath of surrounding
molecules on a Brownian particle is described in terms of a mean-field,
time-dependent stochastic force which is commonly assumed to be white Gaussian
noise. That postulate is compatible with the assumption 
of a short correlation time of fluctuations, much shorter than the time-scale of
the macroscopic
motion and assumes that weak interactions with the bath lead to independent
random variations of the parameter describing the motion.
In more formal, mathematical terms Gaussianity of the state-variable fluctuations
is a consequence of the Central Limit Theorem which states that normalized
sum of independent and identically distributed random variables with finite
variance converges to the Gaussian probability distribution. If, however, 
after random collisions jump lengths are ruled by broad distributions leading to
the divergence of the second moment, the statistics of the process changes
significantly. The existence of the limiting distribution is then guaranteed
by the generalized L\'evy-Gnedenko \cite{gnedenko} limit theorem. According to the latter,
normalized sums of independent, identically distributed ({\it i.i.d}) random variables with infinite variance converge
in distribution to the L\'evy statistics. At the level of the Langevin equation, 
L\'evy noises are generalization of the Brownian motion and describe results of
strong collisions between the test particle and the surrounding environment. In
this sense, they lead to different models of the bath that go beyond a standard 
``close-to-equilibrium'' Gaussian description. 

As documented elsewhere \cite{montroll,shlesinger}, not fully thermalized systems 
or systems driven away from the equilibrium can manifest
interesting physical properties. 
In particular, such systems may exhibit large energy fluctuations with 
probabilities higher than those predicted by the Gaussian statistics.
Non-equilibrated heat reservoir can be thus considered as a source of 
non-Gaussian noises.
Formalisms that give physical background for this phenomena are 
based on the idea of nonextensive thermodynamics,
established on a Tsallis statistics \cite{tsallis}, and the Fractional Fokker Planck 
Equation (FFPE) \cite{klafter}.
The latter (FFPE) have been successfully
applied~\cite{klafter,yuste,sung,sok,sokolov} for
 describing anomalous diffusion processes, for which
the non-local character brought about by the stable L\'evy noise leads to the
replacement of the local spatial derivatives in the diffusion term of the
Fokker-Planck equation by a fractional derivative. 

In contrast to the fractional calculus, the Tsallis statistics approach to
stable, L\'evy-type fluctuations is based on the introduction of a special entropy
 form~\cite{tsallis,tsallis2}
from which examined thermodynamical quantities can be derived~\cite{www}.
The main property of the Tsallis entropy is its nonextensivity, i.e. 
the entropy of the system containing two noninteracting subsystems is
different from the sum of the subsystems' entropies typical for equilibrium Gibssian
ensembles and Gaussian measures.

Various examples of everyday life phenomena and a wide extent of experimental
observations \cite{mandelbrot,bouchaud,montroll} show existence of long-range correlations, disorder, cooperativity
and deviations from the Gaussian statistics, thus suggesting a strong need of study 
more general probability distribution than just Gaussian ones.
In particular, $\alpha$-stable distributions have been observed in anomalous dynamics
and strange kinetics in amorphous semiconductors and glassy systems \cite{sok}.
L\'evy-flight models turn out to be adequate for the description of transport
in heterogeneous catalysis, self-diffusion in micelle systems and analysis of
geophysical data \cite{cates,ott,zu}. Related models have been also
applied in financial modeling and analysis of economic 
time-series~\cite{izydorczyk,mantegnab}.
Among various aspects of the L\'evy-type variables and processes,
 a problem of 
 special interest is 
 numerical generation of stable distributions and simulation of $\alpha$-stable
integrals and stochastic differential
equations~\cite{chamber,janicki1,janicki2,weron1,mantegna,nolan}.
As broadly discussed in literature \cite{janicki1,weron1,nolan}, crucial difficulties 
in solving stochastic differential equations with stable measures
are caused by the noisy term, which allows for larger fluctuations with higher
 probabilities 
than Gaussian distributions.
Numerical methods~\cite{janicki1,izydorczyk} for such equations are more sophisticated 
than for differential equations~\cite{rec}
and for stochastic differential equations with Gaussian noises~\cite{mannella}.
In particular, the nonexistence of variance for stable variables makes the problem much more 
complicated to 
tackle both, numerically~\cite{janicki1, izydorczyk} and
analytically~\cite{garbaczewski,dietlevsen}.
It turns out, however, that with the use of suitable statistical estimation
techniques, computer simulation procedures and numerical discretization methods
it is possible to construct relevant approximations of stochastic integrals
with stable measures as integrators~\cite{janicki1,janicki2}. 

In this communication we study statistical properties of the generic system
 describing passage of a classical particle over a fluctuating potential
 barrier.
 The system is coupled to a non-Gaussian bath modeled by the L\'evy stable noise.
We present numerical results for the mean first
passage time of the particle over the barrier, assuming a linear potential 
subject to Markovian dichotomous fluctuations. 
Our model belongs to the class of ``on--off'' models discussed
in a paper by Doering and Gadoua \cite{doe} and further analyzed by
Bogu\~n\'a \textit{et al.} \cite{boguna}. A distinctive characteristics of
this model is that part of the time the barrier is either switched off (i.e. it becomes 
flat), or the switching is performed between the barrier and a well, so that the
particle can essentially roll rather than climb during these times.
The main difference between the model considered here is a form of
driving, additive fluctuations.
Whereas in the previous papers mainly white Gaussian noises have been considered
 \cite{doe,iwanisz,reiman,dybiec1,dybiec2},
here additive noises are assumed to be $\alpha$-stable~\cite{janicki2},
that might arise from the contact with not fully equilibrated bath.

The problem of resonant activation (RA)~\cite{doe} examined in this paper
is an example of processes manifesting constructive role of noise~\cite{gam}.
The article deals with a modified version of the model proposed by 
Doering and Gadoua~\cite{doe}. Section II presents the model and poses the
problem to be studied.
In Sections III and IV general considerations of L\'evy noises in physical systems 
are addressed
and the problem of underlying thermodynamic interpretation is discussed.
Results of simulations with 
a short note on
 numerical methods applied for generating stable variables and 
 integration
of stochastic differential equations with stable measures
are included in Section V.
The paper is closed with the concluding remarks.

\section{Generic model system}
\begin{figure}[!ht]
\includegraphics[angle=0, width=7.0cm, height=7.0cm]{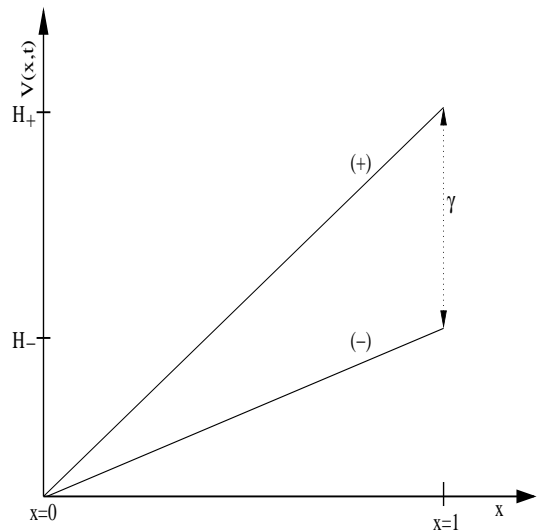}
\caption{\label{plotmodel}A model potential studied in the paper. The barrier
height fluctuates dichotomously between the values $H_{\pm}$.
A particle starts its diffusive motion at a reflecting boundary $x=0$
and continues until the absorption at $x=1$.}
\end{figure}

We consider an overdamped Brownian particle moving in a potential field
between absorbing ($x=1$) and reflecting ($x=0$) boundaries
 (cf.~Fig.~\ref{plotmodel}), in the presence of noise
that modulates the barrier height. 

Time evolution of a state variable $x(t)$ is described in terms of the 
Langevin equation
\bea
\frac{dx}{dt} & =& -V'(x)+g\eta(t)+\zeta(t) \nonumber \\
&= & -V_\pm^{'}(x)+\zeta(t),
\label{lang}
\eea
where prime means differentiation over $x$, $\zeta(t)$ is a white L\'evy process originating 
from the contact with not-fully
equilibrated bath and
$\eta(t)$ stands for a Markovian dichotomous noise of intensity $g$ taking one of
two possible values $\pm 1$. 
Autocorrelation of the dichotomous noise is set to
$\langle (\eta(t)-\langle\eta\rangle)(\eta(t')-
\langle\eta\rangle)\rangle=
\exp(-2\gamma |t-t'|)$. 
For simplicity, throughout the paper a particle mass, a friction coefficient and the
 Boltzman constant are all set to 1. The time-dependent potential 
$V_\pm(x)$ is assumed to be linear with the barrier switching between two configurations
with an average rate $\gamma$ (cf.~Fig.~\ref{plotmodel})
\bea
V_\pm(x)=H_\pm x, \qquad g=\frac{H_{-}-H_{+}}{2}
\eea
Both $\zeta$ and $\eta$ noises are assumed to be statistically independent.

The initial condition for Eq.~(\ref{lang}) is
\be
x(0)=0,
\ee
i.e. initially particle is located at the reflecting boundary with 
equal choices of finding a potential barrier in $(\pm)$ 
configurations
\be
P(H_+,t=0)=P(H_-,t=0)=\frac{1}{2}.
\label{dninit}
\ee
The quantity of interest is the mean first passage time (MFPT)
\be
\mathrm{MFPT}=\int_0^1dx\int_0^\infty[p_-(x,t)+p_+(x,t)]dt,
\ee
i.e., the average time which particle spends in the system 
before it becomes absorbed.
Within the proposed approach, the MFPT is estimated as a first moment 
of the distribution of first passage times (FPT) obtained 
from the ensemble of simulated realizations of the stochastic process in question.
Otherwise, for L\'evy flights described by Eq.~(\ref{lang}), the MFPT may be calculated 
after solving a relevant deterministic fractional Fokker-Planck
equation~\cite{fogedby} (FFPE) for the distribution function
\bea
\frac{\partial p_\pm(x,t)}{\partial t} & = &  \frac{\partial}{\partial x} \left[   \frac{\partial V_\pm(x)}{\partial x}p_\pm(x,t) \right] + D\nabla^\alpha  p_\pm(x,t)  \nonumber \\
 & + & \gamma p_\mp(x,t)-\gamma p_\pm(x,t).
\label{sffpe}
\eea
In the above FFPE $p_\pm(x,t)$ are probability density functions (PDFs) for finding a particle at time $t$ in the vicinity of $x$, 
while potential takes the value $V_\pm(x)$. 
The fractional derivatives $\nabla^\alpha$ is understood in the sense of the Fourier Transform~\cite{jespersen}
\be
\nabla^\alpha=-\int\frac{dk}{2\pi}e^{ikx}|k|^\alpha.
\ee
with $\alpha=2$ corresponding to the standard Brownian diffusion. The coefficient
$D$ denotes the generalized diffusion coefficient with the dimension~\cite{fogedby}
$[D]=cm^{\alpha}sec^{-1}$ and can be related to the parameter $\sigma$ characterizing
the width of the PDF (see below).

\section{L\'evy type variables}

The $\alpha$-stable variables are random variables for which the
the sum of random variables is distributed according to the same
distribution as each variable, i.e.
\be
aX_1+bX_2\stackrel{\mathrm{d}}{=}cX+d,
\label{def}
\ee
where $\stackrel{\mathrm{d}}{=}$ denotes equality in a distribution sense.
Real constants $c,\;d$ in Eq.~(\ref{def}) allow for rescaling and shifting 
of the initial probability distribution.

The characteristic function of the probability distribution that fulfills~(\ref{def}) can be parameterized 
in various ways \cite{nolan}.
In the usually chosen $L_{\alpha,\beta}(\zeta;\sigma,\mu)$ \cite{janicki2,weron1,izydorczyk}
parameterization, 
a characteristic function of the L\'evy type variables is given by 
\bea
 \phi(k) & = & \exp\left[ -\sigma^\alpha|k|^\alpha\left( 1-i\beta\mbox{sign}(k)\tan
\frac{\pi\alpha}{2} \right) +i\mu k\right. \nonumber \\
& -& \left. i\beta k\sigma^\alpha\tan \frac{\pi\alpha}{2}\right],\;\;\; \mbox{for}\;\;\alpha\neq 1, \nonumber \\
 \phi(k) & = & \exp\left[ -\sigma|k|\left( 1+i\beta\frac{2}{\pi}\mbox{sign} (k) \ln|k| \right) \right. \nonumber \\
& + & i\mu k \bigg],\;\;\;\;\;\;\;\;\;\;\;\;\;\;\;\;\; \mbox{for}\;\;\alpha=1,
\label{charakt}
\eea
with
$
\alpha\in(0,2],\;
\beta\in[-1,1],\;
\sigma\in(0,\infty),\;
\mu\in(-\infty,\infty).
$ and $\phi(k)$ defined in Fourier space
\be
\phi(k) = \int d\zeta e^{-ik\zeta} L_{\alpha,\beta}(\zeta;\sigma,\mu)
\ee
Parameter $\alpha$ is called the stability index, $\beta$ describes skewness of the distribution, 
$\sigma$ is responsible for its scaling of and $\mu$ is a location parameter.
The above parameterization (\ref{charakt}) is continuous, in the sense that
\be
\lim\limits_{\alpha\to1}\sigma^\alpha\left(|k|^\alpha\mbox{sign}
(k) -k \right)\tan \frac{\pi\alpha}{2}
=-\frac{2}{\pi}\sigma|k|\mbox{sign} (k) \ln |k|
\ee
for every $\sigma$ and $k$.

Although the PDFs for stable variables $L(\zeta)$ are known to have 
an asymptotic power-law behavior $L(\zeta)\sim |\zeta|^{-(\alpha+1)}$,
analytical expressions corresponding to Eq.~(\ref{charakt}) can be given
only in few cases.
In particular, for $\alpha=2,\beta=0 $, $\zeta$ is a Gaussian variable
with the probability density
\be
L_{2,0}(x;\sigma,\mu)=\frac{1}{2\sigma\sqrt{\pi}}
\exp\left(-\frac{(x-\mu)^2}{4\sigma^2} \right),
\ee
whereas $\alpha=1, \beta=0$ and $\alpha= \frac{1}{2}, \beta=1$ yield Cauchy- 
\be
L_{1,0}(x;\sigma,\mu)=\frac{\sigma}{\pi}\frac{1}{(x-\mu)^2+\sigma^2}.
\label{cauchy}
\ee
and L\'evy--Smirnoff $(x>\mu)$-
\bea
L_{1/2,1}(x;\sigma,\mu)& = & 
\left( \frac{\sigma}{2\pi}
\right)^{\frac{1}{2}}(x-\mu)^{-\frac{3}{2}} \nonumber \\
& \times& \exp\left(-\frac{\sigma}{2(x-\mu)} \right).
\label{smirnoff}
\eea
distributions, respectively.

Generally, for $\beta=\mu=0$ PDFs are asymmetric and for $\beta=\pm1$ and
$\alpha\in(0,1)$ they are totally skewed \cite{janicki2}.

\section{L\'evy noises and Tsallis thermodynamics}

Classical equilibrium thermodynamics can be derived by use of the variational 
principle for 
the Boltzman--Gibbs--Shannon entropy~\cite{reichl}
\be
S[p(x)]=-\int_{-\infty}^\infty p(x)\ln[\sigma p(x)]dx,
\label{gibbs}
\ee
where $p(x)$ is a probability distribution of a given quantity.

For example, $x$ may be associated with a single jump length of a 
free Brownian particle~\cite{tsallis2,prato} 
so that $p(x)$ stands for a single jump length distribution.
Optimizing $S[p(x)]$ with additional constraints imposed on $p(x)$
\be
\int_{-\infty}^\infty p(x)dx=1
\label{norm}
\ee
and 
\be
\int_{-\infty}^\infty x^2p(x)dx=\sigma^2
\ee
leads to a Gaussian distribution $p(x)$ used in the description of a normal Brownian
 diffusion~\cite{klafter}
\be
p(x)=\exp(-b x^2)/Z,
\label{ga}
\ee
Here $Z=\int_{-\infty}^\infty\exp(-bx^2)=\sqrt{\pi/b}$ is a partition function
and
$b$ is a Lagrange multiplier, $b=1/(2\sigma^2)$.

As an extension of the above, a generalized $q$-entropy suggested by
 Tsallis~\cite{tsallis}
\be
S_q[p(x)]= \frac{1-\int_{-\infty}^\infty[\sigma p(x)]^q\frac{dx}{\sigma}}{q-1}
\label{qentropy}
\ee
with a supplementary condition
\be
\int_{-\infty}^\infty x^2[\sigma p(x)]^qd(x/\sigma)=\sigma^2
\label{gpos}
\ee
leads to~\cite{tsallis2} 
\be
p_q(x)=[1-b(1-q)x^2]^{1/(1-q)}/Z_q,
\label{tsa}
\ee
with $Z_q=\int_{-\infty}^\infty [1-b(1-q)x^2]^{1/(1-q)} dx$.
In the limit of $q\rightarrow 1, p_q(x)$ reproduces Eq.~(\ref{ga}) and the 
norm-constraint
can be satisfied only for $q<3$. $N$-folded distribution $p_q(x,N)$ approximating probability
distribution for macroscopic diffusion is determined by the general central limit theorem.
Consequently, for $q<5/3$, the asymptotic ($N\rightarrow\infty$) distribution is 
Gaussian
whereas for $q>5/3$ it approaches the L\'evy statistics
 \cite{tsallis,tsallis2}
\be
p_q(x,N)\approx L_{\alpha}((\sqrt{b}/N^{1/\alpha})x)
\ee
with the stability index $\alpha$ given by
\be
\alpha=\frac{3-q}{q-1}\;\;(5/3<q<3).
\ee
The variational principle 
can be also applied to the energy distribution~\cite{tsallis3} which subject to the 
generalized condition
\be
\sum_i[p(\varepsilon_i)]^q\varepsilon_i=U,
\label{energy2}
\ee
results in 
\be
p_q(\varepsilon_i)\propto\left[ 1-b(1-q)\varepsilon_i\right]^{1/(1-q)},
\label{endist2}
\ee
Alternatively, expression Eq.~(\ref{endist2}) can be derived by assuming subordination of
the Gibssian exponential energy statistics to the statistics of fluctuations in 
 temperature $T$ (and hence in the parameter $b=1/T$)~\cite{wilk}
 \be
 p_q(\varepsilon_i)\propto\int^{\infty}_0\exp(-b{\varepsilon})f(b)db
 \ee
 In particular, a gamma-distributed temperature leads to a L\'evy
 distribution $p_q(\varepsilon)=L_{\alpha}(b\varepsilon)$ with the stability index given by the
 relative variance of fluctuations in $b$ ~\cite{wilk}.
This observation shows a possible link of the L\'evy distributions to models of
nonequilibrated baths. 

\section{Physical interpretation}

By definition, in the case of static barrier height and with the noise $\zeta$
 uncorrelated at different times and obeying the L\'evy statistics, the overdamped Langevin equation
Eq.~(\ref{lang}) describes L\'evy flights \cite{fogedby,jespersen} in a constant force
field. For long times, the trajectory $x(t)$ behaves as
\be
x(t)\approx H_0t+\int^t_0 ds\zeta(s)
\label{owlang}
\ee
and 
yields the L\'evy stable distribution~\cite{nolan,jespersen} in the position of the
 particle. In consequence, if the first moment exists, i.e. for $1<\alpha\le 2$, 
 mean value of $x(t)$ grows linearly with time, $\langle x(t) \rangle=H_0t$ whereas the mean square
 displacement becomes $\langle [x(t)- \langle x(t) \rangle ]^2 \rangle =2D=2\sigma^2$ only for $\alpha=2$ when the
 finite second moment of $L(\zeta)$ exists.
 Thus the generalized Einstein relation connecting the first moment in the presence
 of constant force $H_0$ to the second moment in the absence of force
 $\langle x(t) \rangle_{ H_0}=H_0 \langle x^2(t) \rangle_0$ is recovered only in the Brownian limit $\alpha=2$
 with the noise amplitude $\sigma$ related to the diffusion coefficient
 $\sigma=D^{1/2}$. Obviously, for $\zeta$ noises with diverging mean square
 displacement, the classical fluctuation dissipation theorem is violated
 and the Einstein relation does not longer hold \cite{sokolov,barkai,jespersen}. 
 
 When analyzed from the perspective of the continuous time random walks (CTRW), L\'evy flights
 characterize walks with a Poisson waiting time and a L\'evy distribution of the jump length.
 The scaling nature of the jump length PDF leads then to a clustering of the L\'evy flights
 visible {\it via} interruption of the local motion by occasional long sojourns on all length scales
  \cite{klafter}. This fractal character of the L\'evy flight-trajectory can be contrasted with subdiffusive
 CTRW \cite{klafter,sokolov} that fills the two-dimensional space completely 
 and features no clusters.
 In such a case, however, the time intervals between consecutive steps are 
 governed by the power-law
 waiting time distributions that lead to a sublinear 
 dependence $\langle x(t) \rangle \approx t^{\nu}$ with $\nu$
 denoting the power-law index of the waiting time PDF. Therefore, subdiffusive CTRW in a constant
 external force acting along the $x$ direction perfectly satisfies the fluctuation-dissipation theorem 
 \cite{klafter,sokolov} which holds also in the corresponding fractional Fokker-Planck equation
framework. 

 Diverging mean-square displacement cannot be valid for a particle with non-diverging mass. In fact, for
 massive particles, a finite velocity of propagation exists making very long instantaneous jumps
 impossible. For that reason, the dilemma of diverging mean-square displacement in the L\'evy flight can
 be overcome by CTRW version of L\'evy walks with a suitable time cost penalizing long jumps.
 Nevertheless, in many physical systems of interest diffusion of state variable $x(t)$ with diverging
 second moment does not violate physical principles and is a legitimate way of physical modeling
  \cite{cates,ott,silbey,may}.

\section{Stochastic differential equations with stable noises}

Instead of solving~(\ref{sffpe}) the corresponding Langevin equation~(\ref{lang}) 
can be simulated by use of the appropriate numerical methods.
Position of the particle is then obtained by direct integration of Eq.~(\ref{lang})
\bea
x(t) & =& -\int_{t_0}^{t}\left[ V'(x(s))-g\eta(s)\right] ds \nonumber \\
 & + & \int_{t_0}^{t} dL_{\alpha,\beta}(s) \nonumber \\
&= &
-\int_{t_0}^{t}V_\pm^{'}(x(s))ds+\int_{t_0}^{t}dL_{\alpha,\beta}(s).
\label{lcalka}
\eea
In general~\cite{janicki2,izydorczyk,dietlevsen}, the $L_{\alpha,\beta}$
 measure in Eq.~(\ref{lcalka}) can be approximated by
\bea
\int_{t_0}^{t}f(s)dL_{\alpha,\beta}(s) & \approx & 
\sum\limits_{i=0}^{N-1}f(i\Delta s)M_{\alpha,\beta}\left([i\Delta s,(i+1)\Delta s)\right) \nonumber \\
& \stackrel{\mathrm{d}}{=} & \sum\limits_{i=0}^{N-1}f(i\Delta s)\Delta
s^{1/\alpha}\varsigma_i,
\eea
where $\varsigma_i$ is distributed with the PDF
$L_{\alpha,\beta}(\varsigma;\sigma=1,\mu=0)$, $N\Delta s=t-t_0$ and $M_{\alpha,\beta}([i\Delta s,(i+1)\Delta s))$ is the measure of the interval $[i\Delta s,(i+1)\Delta s)$.

Random variables $\varsigma$ corresponding to the characteristic function
 (\ref{charakt}) can be generated using the Janicki--Weron algorithm
 \cite{janicki1,izydorczyk,weron1}.
For $\alpha\neq1$ their representation is 
\bea
\varsigma & = & D_{\alpha,\beta,\sigma} \frac{\sin(\alpha(V+C_{\alpha,\beta})) }{
(\cos(V))^{\frac{1}{\alpha}}} \nonumber \\
& \times & \left[
\frac{\cos(V-\alpha(V+C_{\alpha,\beta}))}{W}
\right]^{\frac{1-\alpha}{\alpha}}+B_{\alpha,\beta,\sigma,\mu}.
\label{recipe1}
\eea
with constants $B,C,D$ given by
\be
B_{\alpha,\beta,\sigma,\mu}=\mu-\beta\sigma^\alpha\tan(\frac{\pi\alpha}{2}),
\ee
\be
C_{\alpha,\beta}=\frac{\arctan\left(\beta\tan(\frac{\pi\alpha}{2})\right)}{
\alpha},
\ee
\be
D_{\alpha,\beta,\sigma}=\sigma\left[ \cos\left(
\arctan\left(\beta\tan(\frac{\pi\alpha}{2})\right) \right) \right]^{-\frac{1}{\alpha}}.
\ee
For $\alpha=1$, $\varsigma$ can be obtained from the formula
\bea
\varsigma & = & \frac{2\sigma}{\pi} \left[ (\frac{\pi}{2}+\beta V)\tan(V) -\beta\ln
\left( \frac{\frac{\pi}{2}W\cos(V)}{\frac{\pi}{2}+\beta V}
\right) \right] \nonumber \\
& + & B_{1,\beta,\sigma,\mu}
\label{recipe2}
\eea
with
\be
B_{1,\beta,\sigma,\mu}=\mu+\frac{2}{\pi}\beta\sigma\ln(\sigma).
\ee
In the above equations $V$ and $W$ are independent random variables, such that
$V$ is uniformly distributed in the interval
$(-\frac{\pi}{2},\frac{\pi}{2})$ and
$W$ is exponentially distributed with a unit mean \cite{janicki2,izydorczyk,weron1}.

The problem described by Eq.~(\ref{lang}) and corresponding
Eq.~(\ref{lcalka})
for the stability index $\alpha=2$ is a well known case of resonant activation 
resolved in the series of papers
 \cite{doe,dybiec1,dybiec2}.
Accordingly, for $\alpha=2$ integration in Eq.~(\ref{lcalka}) is performed 
with respect to a Gaussian measure and standard numerical methods can be
applied \cite{sobczyk} to solve the Langevin equation under study. Otherwise, the ordinary forward (or backward) in time Fokker
 Planck Equation~\cite{risken,gardiner,kam} for the PDF can be derived
 in a closed form, from which the MFPT can be easily calculated. Also, for the
 Gaussian fluctuations theoretical background is provided by the standard thermodynamics
  \cite{reichl} where parameter $\sigma$ can be related to the intensity of the
 thermal fluctuations imposed on a physical mode $x$.

Within these studies, besides the driving Gaussian fluctuations,
two other kinds of $\alpha$-stable noises have been
considered, namely 
the Cauchy noise ($\alpha=1,\;\beta=0$) and the Smirnoff noise
 ($\alpha=0.5,\;\beta=1$) with intensities $\sigma=0.5$, $\sigma=1/\sqrt{2}$ and $\sigma=1$.
The value of the location parameter $\mu$ has been arbitrary set to 0.

The choice of the $\sigma$, which scales the distribution width and varies the
intensity of fluctuations,
 corresponds, although not in a self-transparent way,
 to the change of system temperature. However, the problem of definition of 
 the system temperature is more subtle here than in
a presence of a standard bath \cite{tsallis2,annunziato,wilk,beck,cohen} because the system 
is not in the state of equilibrium. 

L\'evy type variables 
have been generated by use of recipes given by Eq.~(\ref{recipe1}) and~(\ref{recipe2}).
The trajectory $x(t)$ has been generated according to~(\ref{lcalka}) for
 various noise realizations. Its evaluation proceeded
 till the time $t'$ when $x(t')\ge 1$ for the first time.
The resulting distribution of first passage times (FPT) has been further used to
evaluate MFPT.

\begin{figure*}[!ht]
\includegraphics[angle=0, width=16.5cm, height=17.0cm]{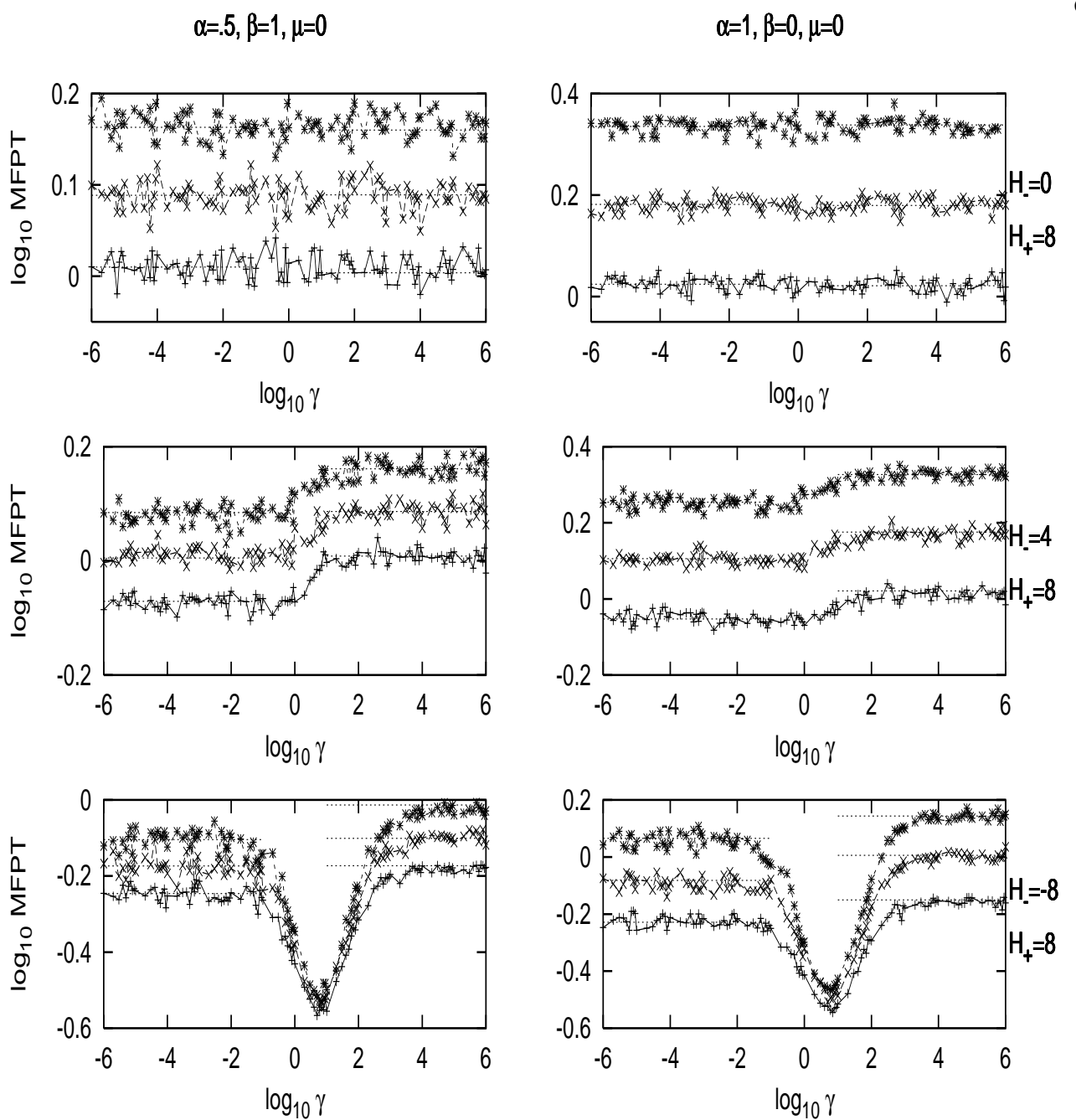}
\caption{\label{plotl}The $\mathrm{MFPT}(\gamma)$ as a function of barrier fluctuation rate $\gamma$ for linear
potentials with $H_+=8,\;H_-=0$ (upper panel), $H_+=8,\;H_-=4$ (middle panel) $H_+=8,\;H_-=-8$ (lower panel)
are plotted along with asymptotic lines.
The additive noise is of the L\'evy--Smirnoff type (left panel)
 ($\alpha=0.5,\;\;\beta=1,\;\;\mu=0$) and of the
 Cauchy type (right panel) ($\alpha=1,\;\;\beta=0,\;\;\mu=0$), respectively.
Different symbols correspond to varying intensity of fluctuations:
 $\sigma=.5$ ($+$), $\sigma=1/\sqrt{2}$ ($\times$) and $\sigma=1$ ($\ast$).
Numerical results have been obtained by use of Eq.~(\ref{lcalka}) with the time step $dt=10^{-5}$. 
Noises were generated according to Eqs.~(\ref{recipe1}) and (\ref{recipe2}).
In order to gain sufficient statistics averaging of Eq.~(\ref{lcalka}) over $10^3$ realizations
has been performed.
Asymptotic lines have been evaluated for reference static potential profiles with the time step $dt=10^{-5}$ and averaged over $10^4$ realizations.
Error bars have been
estimated by use of standard techniques for the MC data 
analysis \cite{newman}. They represent standard deviation from the mean and 
 remain within the symbol size. 
Lines have been drawn to guide the eye.
}
\end{figure*}

The linear potential $V_\pm={H_\pm}x$ has been dichotomously alternating 
between different values of $H_\pm$.
Our simulations have been obtained for $H_\pm=\pm8$ (cf. Fig.~\ref{plotl}~(lower panel)),
 $H_-=0,\;H_+=8$ (cf. Fig.~\ref{plotl}~(upper panel)) and $H_-=4,\;H_+=8$ (cf. Fig.~\ref{plotl}~(middle panel)).

Asymptotic lines plotted in Fig.~\ref{plotl} have been calculated numerically by use of
the Monte Carlo method \cite{newman}.
As expected, typical behavior \cite{doe,boguna,dybiec1} has been recovered
for $\gamma\to0$
\be
\mathrm{MFPT}(\gamma\to 0)=\frac{1}{2} \left(\mathrm{MFPT}(H_-)+\mathrm{MFPT}(H_+) \right),
\label{aszero}
\ee
and for $\gamma\to\infty$
\be
\mathrm{MFPT}(\gamma\to \infty)=\mathrm{MFPT}\left(\frac{1}{2}(H_-+H_+)\right).
\label{asinf}
\ee
Fig.~\ref{plotl} presents results of simulations averaged over $10^3$
 realizations with a time-step $\Delta t=10^{-5}$.
For calculation of the asymptotic lines (Eqs.~(\ref{aszero}) and 
 (\ref{asinf}) ) $\Delta t=10^{-5}$ and averaging over $10^4$ realizations 
has been performed.

For the barrier switching between $H_\pm=\pm8$, the phenomenon of resonant activation is clearly visible cf. Fig.~\ref{plotl}~(lower panel).
By comparison to the results obtained for a motion of a particle subject to 
the additive white Gaussian noise~\cite{doe,boguna,iwanisz,reiman,dybiec1}, 
the value of the MFPT is smaller and resonant activation is observed at lower
frequencies
 $\gamma$. 
Moreover, for $H_\pm=\pm8$, asymptotic values of MFPT estimated for 
$\gamma\to\infty$ and $\gamma\to0$ are higher (lower)
 than in the corresponding white Gaussian noise case~\cite{doe,boguna,dybiec1}.
 
For other cases under consideration, i.e. for $H_-=0,\;H_+=8$ and 
$H_-=4,\;H_+=8$, the resonant activation has not been observed (cf. Fig.~\ref{plotl}~(upper and middle panel)).
However, values of reported MFPTs for the system driven by non-Gaussian L\'evy noises
 are always significantly smaller than in the case of Gaussian-noise driving.
It is caused by the fact that the Cauchy and L\'evy--Smirnoff noises allow for
 higher values of driving fluctuations in the Langevin equation (\ref{lang}).

As it can be inferred from Fig.~\ref{plotl}, values of the MFPT
for the Cauchy noise are higher than the corresponding times for the L\'evy--Smirnoff noise, compare the left and the right panel of the Fig.~\ref{plotl}.
This observation can be explained by the difference between both statistics: 
for the given sets of parameters, the L\'evy--Smirnoff distribution is more ``heavily tailed'' than the Cauchy distribution.

The procedure applied for numerical integration of~(\ref{lcalka}) is valid for every allowed value of $\alpha$.
In particular, the case with $\alpha=2$ has been investigated along these lines in order to test
the numerical approach adopted in simulation of Eq.~(\ref{lang}). The test concluded a perfect
agreement of numerically simulated results with formerly solved Gaussian cases of RA
 \cite{doe,reiman,dybiec1,dybiec2}.

\section{Summary}
We have considered a thermally activated
process that occurs in a system coupled to a non-Gaussian noise
source introduced by a not fully equilibrated thermal bath. Another external
stochastic process is assumed to be responsible for dichotomous fluctuations of the potential
barrier which has
been modeled by the linear function with a varying slope.

In comparison to the Gaussian case~\cite{dybiec2}, when the RA phenomenon
 has been observed for all barrier setups ($H_\pm=\pm8,\;H_+=8,\;H_-=0,\;H_+=8,\;H_-=4$)
 analyzed in this study, non-Gaussian
additive stable noises produce resonant activation observable only for the $H_\pm=\pm8$ case.
Resulting \mbox{MFPTs} are significantly smaller than those obtained for the Gaussian
source of fluctuations. The effect is due to higher probabilities of the extreme events (large
fluctuations) allowed by the L\'evy statistics.
Similarly to the Gaussian-bath case, a typical asymptotic behavior of the MFPT has been 
recovered for large and small
frequencies of the dichotomous noise:
for small $\gamma$ MFPT tends to the average MFPTs for the both barrier configurations,
while for large $\gamma$ it becomes equal to the MFPT over the average potential barrier
 \cite{doe,iwanisz,reiman}.
 
These observations have several implications in relation to chemical kinetics
in conformationally varying media \cite{ewa,schmid,berez} where flipping barriers
separating reactants' and products' basins may be due to a dynamic isomerization
of the activated complex. 
The effects of fluctuations in force or potential on the diffusive process have
been also extensively studied in the context of motor proteins \cite{astumian}.
Under typical conditions, i.e. with a Gaussian  additive thermal noise and with
a flipping barrier height, the RA phenomenon is registered in all those realms as a maximal flux of
 particles or a maximum  reaction rate. In contrary to this finding, the analogous systems
 influenced by the presence of non-Gaussian additive noises  exhibit RA only for potentials
 switching between the barrier and the well (Fig.~\ref{plotl}, lower panel). Otherwise,
 the  kinetics is fairly insensitive to 
 the flipping rate of an erecting barrier (Fig.~\ref{plotl}, upper panel) or becomes
 slowed down (Fig.~\ref{plotl}, middle panel)  when the frequency of switching between 
a high and a low barrier becomes large.

In addition, physical interpretation of the L\'evy noisy term in equation (\ref{lang}) was discussed in some detail.
In general, descriptions provided by the FFPE approach and the Tsallis statistics 
are not in a full agreement \cite{fogedby,klafter,jespersen}.
In particular, the FFPE description of L\'evy flights in a harmonic potential
 \cite{jespersen,zanette}
produces different stationary PDF than the one predicted by the entropy Eq.~(\ref{qentropy}).
Nevertheless, the Tsallis formalism of nonextensive statistical mechanics (or the ``superstatistics'' 
concept, in general \cite{beck,cohen}) offers 
an intriguing interpretation of non-Gibssian ensembles that can model not fully equilibrated
baths. Especially, for nonequilbrium systems composed of regions that exhibit spatiotemporal
fluctuations of an intensive quantity (like pressure, chemical potential, inverse temperature or
the energy dissipation rate \cite{beck}) generalized statistics may emerge in consequence
of a statistical subordination of the intensive variable to those fluctuations.

The situation in which resulting probability distributions are not Gaussian 
are frequent in various physical situations.
The phenomena of special interest are random walks that may lead 
to L\'evy distributions~\cite{klafter} and hence provide a possible realization of
 L\'evy noise sources.

There are several candidates for such a noise origin to be taken into account.
Obviously, all noise sources need to be infinite to allow, with a nonzero probability,
infinite fluctuations.
Among possible models of chaotic nonequilibrium baths perhaps the most natural is
 a model of the fluid--like bath~\cite{baranger} abruptly perturbed with a local 
heating to a very high temperature. 
Gradual spreading of energy progresses until a new equilibrium state is reached
and for some finite periods of time, the probing system that experiences local change of the
 bath temperature will be governed by non-Boltzman statistics \cite{cohen,baranger,wilk}.
 This and similar approaches validate then use of the L\'evy-type statistics for description of
 not-fully equilibrated baths. As known from other similar studies \cite{dietlevsen},
the L\'evy-type structure of the noise imposed on a double-well potential can 
affect profoundly noise induced jumping between
metastable states and result in stationary PDF deviating from the usual Gibbs distribution.

\begin{acknowledgments}
Authors acknowledge stimulating and inspiring discussions with P.~F.~G\'ora.
\end{acknowledgments}



\end{document}